\documentclass[11pt]{amsart}
\usepackage{amsmath, amssymb}
\usepackage[colorlinks=true, linkcolor={red}, 
   citecolor={blue},urlcolor={magenta}]{hyperref}
\usepackage[margin=3cm]{geometry}
\usepackage{graphicx}

\def\hotscan{{\tt hot\_scan}}
\newcommand{\Prb}{\mathbb P}
\newcommand{\Arv}{\text{AID}^{rv}}
\newcommand{\Amm}{\text{AID}^{\text{\sc --/--}}}
\newcommand{\Bmm}{\text{53BP1}^{\text{\sc --/--}}}
\newcommand{\BArv}{\text{53BP1}^{\text{\sc --/--}}\text{AID}^{rv}}
\newcommand{\BAmm}{\text{53BP1}^{\text{\sc --/--}}\text{AID}^{\text{\sc --/--}}}

\title[Chromosome-wide translocation hotspots]{%
  Identification of cromosomal translocation hotspots via scan 
  statistics
 }
\date{October 9, 2013}

\begin{document}

\maketitle

\centerline{\sc Israel T. Silva$^{*,\sharp,}$\footnote{To whom
  correspondence should be addressed}$^,$\footnote{This is an
    un-refereed author version of an article  submitted for
    publication in \emph{Bioinformatics}}, 
   Rafael A. Rosales$^\dag$, 
   Adriano J. Holanda$^\dag$,}
\centerline{\sc Michel C. Nussenzweig$^*$ and Mila Jankovic$^*$} 
\medskip
  \centerline{\small $^*$Laboratory of Molecular Immunology, The
    Rockefeller University}
  \centerline{\small 1230 York Avenue, New York, NY 10065}
\medskip
   \centerline{\small $^\dag$Departamento de Computa\c{c}\~ao e
     Matem\'atica, Universidade de S\~ao Paulo} 
   \centerline{\small Av. Bandeirantes, 3900, Ribeir\~ao Preto,
     CEP 14049-901, SP, Brazil}
\medskip
   \centerline{\small $^\sharp$National Institute of Science and
     Technology in Stem Cell and Cell Therapy} 
   \centerline{\small and Center for Cell Based Therapy} 
   \centerline{\small Rua Cat\~ao Roxo, 2501, Ribeir\~ao Preto, CEP
     14051-140, SP, Brazil}
\medskip

%
% -----------------------------------------------------------
%
\begin{abstract}
  {\bf Motivation:}
  The detection of genomic regions unusually rich in a given pattern
  is an important undertaking in the analysis of next generation
  sequencing data.  Recent studies of chromosomal translocations in
  activated B lymphocytes have identified regions that are frequently
  translocated to c-myc oncogene. A quantitative method for the
  identification of translocation hotspots was crucial to this
  study. Here we improve this analysis by using a simple probabilistic
  model and the framework provided by scan statistics to define the
  number and location of translocation breakpoint hotspots. A key
  feature of our method is that it provides a global chromosome-wide
  significance level to clustering, as opposed to previous methods
  based on local criteria. Whilst being motivated by a specific
  application, the detection of unusual clusters is a widespread
  problem in bioinformatics. We expect our method to be useful in the
  analysis of data from other experimental approaches such as of
  ChIP-seq and 4C-seq.\\
  {\bf Results:} 
  The analysis of translocations from B lymphocytes with the method
  described here reveals the presence of longer hotspots when compared
  to those defined previously. Further, we show that the hotspot size
  changes quite substantially in the absence of DNA repair protein
  53BP1. When 53BP1 deficiency is combined with overexpression of
  activation induced cytidine deaminase (AID) the hotspot length
  increases even further. These changes are not detected by previous
  methods that use local significance criteria for clustering. Our
  method is also able to identify several exclusive translocation
  hotspots located in genes of known tumor supressors.\\ 
  {\bf Availability:} The detection of translocation hotspots is
  done with \hotscan, a program implemented in R and Perl. Source
  code and documentation are freely available for download at
  \href{https://github.com/itojal/hot\_scan}
  {\texttt{https://github.com/itojal/hot\_scan}}.\\
  {\bf Contact:}
  \href{mailto:isilva@rockefeller.edu}{\tt isilva@rockefeller.edu}%,
  %\href{mailto:itojal@gmail.com}{\tt itojal@gmail.com}.
\end{abstract}

\section{Introduction}
The identification of genomic regions that are unusually rich in a
given pattern is a recurring problem in bioinformatics, quite
widespread in the analysis of data generated by deep-sequencing. An
example of this is the detection of regions with an unlikely high
clustering of chromosomal translocation breakpoints (\cite{Hasan10,
Jankovic13, Klein11}). Recurrent chromosomal translocations are
associated with hematopoietic malignancies such as leukemia and
lymphoma and with some sarcomas and carcinomas (\cite{Kuppers05,
Kuppers01, ZweiNusse10, Rabbitts09, Kumar08}). There is growing
evidence that translocations are not random. Among basic determinants
of these events are the existence of chromosome territories, active
transcription and most prominently targeted DNA damage
(\cite{Klein11, Chiarle11, Hakim12}). DNA double
strand breaks (DSBs) are necessary intermediates in chromosome
rearrangements and they occur in the cell during normal metabolic
processes, and can be induced by genotoxic agents or during
physiological DNA recombination in lymphocytes. The majority of human
lymphomas are of mature B cell origin and many of them carry balanced
chromosomal translocations that involve immunoglobulin genes
(\cite{Kuppers05}). This susceptibility is most likely dependent on
Activation-induced cytidine deaminase (AID), the B lymphocyte specific
enzyme that initiates class switch recombination (CSR) and somatic
hypermutation (SHM), two processes nevessary for antibody
diversification (\cite{Revy00, Ramiro06}). AID initiates SHM and
CSR by deaminating cytosines in immunoglobulin genes during stalled
transcription (\cite{Chaudhuri04, Storb07, Pavri10}). Several DNA
repair pathways process the resulting U:G mismatch to introduce
mutations or produce targeted DSB (\cite{DiNoia07,
Stavnezer08}). Besides being targeted to immunoglobulin genes, AID
targets a large number of non-immunoglobulin genes (\cite{Liu08,
Pavri10, Yamane11}).  AID induced DSBs are recognised by DNA damage
response (DDR) proteins and repaired by non-homologous end joining
(NHEJ), a process that can fail and result in chromosomal
translocations (\cite{ZweiNusse10, Zhang10, Gostissa11}).
Libraries of AID dependent translocations from primary B cells
revealed many discrete sites throughout the genome that are targeted
by AID. Some of these targets are known translocation partners
identified in mature B cell lymphomas \cite{Klein11,
Chiarle11}. Mutations in the components of DNA repair pathways that
process AID induced breaks can lead to defective CSR and the most
severe defect is documented in 53BP1 deficient B cells. 53BP1 is a DNA
repair protein that regulates DSB processing and is required for
genomic stability. It does so by facilitation distal DSB joining and
by protecting DNA ends from resection (\cite{Difilippantonio08,
  Bunting10, Bothmer11}). The landscape of AID induced 
translocations in 53BP1 deficient B cells is different from the one
found in wild type cells. Deep sequencing of translocation capture
libraries from primary 53BP1 deficient B lymphocytes has shown that
the profile of translocation hotspots changes most likely due to
increased DNA end resection (\cite{Jankovic13}).

A quantitative method to determine the clustering of translocations
was essential for the analysis of chromosomal rearrangements in
\cite{Klein11} and \cite{Jankovic13}. Translocation hotspots were
determined by using a technique similar to that used to define the
coordinates of enriched protein binding regions in ChIP-seq
experiments. A translocation cluster is defined by concatenating
closely spaced adjacent breakpoints and its significance is then
determined by using a test based on the negative binomial
distribution. This method assumes that the observed breakpoints are a
realisation of a Bernoulli process. By taking advantage of this model
here we consider a different approach for the detection of hotspots
based on the use of scan statistics, \cite{Glaz01, Balakrishnan}. The
scan statistic is particularly suited in the current setting because
if provides a genome-wide significance level to breakpoint 
clustering. Using our method we are able to show that translocation
hotspots induced by AID in activated B lymphocytes are longer than
those previously identified by the local method. Furthermore, our
method shows that long hotspots are more frequent in the absence of
53BP1. The frequency of the long hotspots is further increased if AID
is overexpressed in 53BP1 deficient B cells. We also discover a set of
hotspots exclusively found by the scan statistic and discuss the
potential biological relevance of our findings.

\section{Methods}
Several methods have been developed to detect clustering of 
events when the observations arise from a spatial or temporal point
process. This section describes the application of the scan statistic
for the detection of genomic regions with a particularly high density
of translocations. We also describe explicitly a technique previously
used by \cite{Klein11}, and \cite{Jankovic13}, which attributes
a local significance level to a hotspot. A mayor difference between
these methods is that the scan statistic provides a global
chromosome-wide  significance level to clustering.

\subsection{Scan statistic approximations}\label{subsec:ss}
We model the occurrence of translocation breakpoints in a
chromosome of length $N$ as realisation of an independent and
identically distributed sequence of 0-1 Bernoulli random variables,
$\xi_1, \ldots, \xi_N$ with $\Prb\{\xi_j=1\} = p$ and $\Prb\{\xi_j=0\}
= 1-p$, $1\leq j \leq N$ for $0 <p <1$. The event $\{\xi_j = 1\}$
occurs if there is a translocation breakpoint at the $j$-th base. We
refer hereafter to this model as the global chromosome null hypothesis,
$H_0$. Let $m$ be a positive integer and then let 
$$Y_i = 
\sum_{j = i}^{i+m-1} \xi_j, \quad 1 \leq i \leq N-m+1$$
be the running number of successes in a window of width $m$. The scan
statistic $S_m$ is defined as the maximum number of successes within
any of the $N-m+1$ consecutive windows,
\[
  S_m = \max_{1\leq i \leq N-m+1}\{Y_i\}.
\]
The significance of a cluster of translocation events in a window of
width $m$ can be assessed by the probability of the tail event
$\{S_m\geq y\}$. Small probabilities for this event indicate
departures for the Bernoulli model consistent with $H_0$ and could
therefore be used to detect hotspots. A considerable effort has been
made in order to derive the distribution of $S_m$ under $H_0$. Still
in this simple case its form has remained elusive. Several
approximations and asymptotics for the distribution of the scan
statistic have been derived under the Bernoulli model, particularly
when the number of observed events in $N$ trials, $a$, are
known. Following \cite{Naus74,Glaz01}, the conditional probability
$\Prb\{S_m \geq y \mid a\}$ may be approximated by the function
\begin{align}
  \notag
  \varphi(y) 
  = 
  &2\sum_{i=y}^a H(i; a, m, N)\\
  &+ \Big(y\frac{N}{m} - a - 1\Big) H(y; a, m, N), \label{eqn:Naus} 
\end{align}
with $H(y; a, m, N)$ as the hypergeometric distribution,
\[
  H(y; a, m, N) = {m\choose y}{N-m\choose a-y}\Big/{N\choose a}.  
\]

Although the expression in (\ref{eqn:Naus}) already provides a method
to quantify the significance of a cluster, the following asymptotic
version avoids the use of the hypergeometric distribution and allows
for an efficient implementation. For sufficiently large $m$ and $N$,
the function in (\ref{eqn:Naus}) may be approximated by
\begin{equation}
 \label{eqn:prepvalue}
   \varphi(y) 
  \approx 
    2 \sum_{i=y}^{a} b(i; a, \theta)  + \Big(y \frac{N}{m} 
   -a -1\Big)b(y; a, \theta)
\end{equation}
where $b(y; a, \theta)$ denotes the Binomial distribution for $a$
trials and success probability $0 < \theta = m/N < 1$. This
approximation is ensured by weak convergence of the hypergeometric
distribution towards the binomial law for large populations and
becomes very accurate in the current application where $N > 1\times
10^6$ and $m \geq 500$.  Furthermore, for large values of $a$, namely
for $1000 \leq a \leq 10000$ as is the case for most chromosomes in
the data sets considered here, the summation in (\ref{eqn:prepvalue})
may be evaluated as
\[
  \sum_{i=y}^a b(i; a, \theta) =b(y; a, \theta)\ 
    _2F_1\Big(1, y-a; 1+y; \frac{\theta}{\theta-1}\Big),
\]
with $\ _2F_1$ as Gauss hypergeometric function, that is, for
$|\mathit z| < 1$ and $n_1$ $n_2$, $n_3 \in \mathbb Z$, 
\[
   _2F_1(n_1, n_2; n_3; \mathit z) = \sum_{i=0}^\infty
   \frac{(n_1)_i (n_2)_i}{(n_3)_i}
   \frac{\mathit z^i}{i!}
\]
and $(n)_i$ as the $i$th Pochhammer symbol of $n \in \mathbb
N$, i. e. $(n)_i = n(n+1)\cdots(n+i-1)$. Note that the second argument
of $ _2F_1$ is always negative or zero because $y \leq
a$. The series defining $\ _2F_1$ is thus finite. With this
simplification the desired $p$-value
\[
  \widehat p = \Prb\{S_m \geq y\mid a\}
\] 
is approximately 
\begin{equation}
  \widehat p \approx
  b(y; a, \theta)\Big\{
    2\ _2F_1\Big(1, y-a; 1+y; \frac{\theta}{\theta-1}\Big)
    + \Big(y \frac{N}{m} -a -1\Big)\Big\}.\label{eqn:THEpvalue}
\end{equation}

We observe that (\ref{eqn:prepvalue}) is the approximation for the
probability of $\{S_m \geq y \mid a\}$ described by
\cite{WallNeff87} in the well known continuous case, namely when $N$
points are drawn uniformly from $[0, 1)$ and $S_m$ is the largest
number of points to be found in any subinterval of $[0, 1)$ of length
$m$. Despite of the existence of several other approximations for the
probability of $\{S_m \geq y\mid a\}$, \cite{Glaz89} observes that
this approximation is quite precise when the right side of
(\ref{eqn:prepvalue}) is less or equal to 0.01 and recommends its use
in this regime.

The detection of chromosomal translocation breakpoints via scan
statistics has also been considered by several authors in the analysis
of leukemias, see \cite{Berger13, Busch07, Hasan10, Reiter03,
Wiemels02}.  These analysis are based on the method described by
\cite{Segal02}, by following a large deviation approximation for
the probability of $\{S_m \geq y\}$ described in
\cite{Loader91}. Although being derived by using rather different
arguments, \cite{Naus04} observe that this approximation and the
one in (\ref{eqn:Naus}) produce similar results.

\subsection{Hotspots}\label{sec:SS_HS}
The method outlined in Section~\ref{subsec:ss} provides a
significance test for the existence of hotspots, still their actual
number and location have to be determined. Here we describe a method
to infer the coordinates of these events.

A chromosome-wide scan with a window of width $m$ leads to the
consideration of the following sequence of local null hypotheses. 
For $i =
1, \ldots, N - m + 1$, 
\begin{equation*}
   H_{0, i}:\  \xi_i, \ldots, \xi_{i+ m} \quad 
   \text{are i.i.d. Bernoulli}(p) \text{ random variables}.
\end{equation*}
Let $y_{i}$ be the observed number of translocation events in the
$i$-th window, $w_i$. The hypothesis $H_{0,i}$ is rejected at a
prescribed level $\alpha_H$ if $\widehat p_{i} = \Prb\{S_{m} \geq y_i\}
\leq \alpha_H$, with $\widehat p_i$ computed according to
(\ref{eqn:THEpvalue}). This procedure partitions the chromosome into
two regions
\[
   \mathfrak B = \big\{w_{i}: \widehat p_{i} \leq \alpha_H, i \in
            [1,N-m+1]\big\}  
 \quad\text{and}\quad 
   \mathfrak B^C.
\]
The connected components of $\mathfrak B$ are prospective hotspot
candidates formed by one or more scanning windows of width $m$. To
actually account for the bias involved while considering the
simultaneous rejection of the multiple hypotheses involved in a given
component of $\mathfrak B$, we adjusted the corresponding $p$-values
by using the Benjamini-Yekutieli correction, \cite{BenYak01}. This
correction accounts for the possibility of having a positive
dependence structure among the considered set of hypotheses from
overlapping scan windows. A similar control of the false discovery
rate associated with the large number tests produced by the scan
statistic has previously been considered while scanning for clusters
in random fields by \cite{Perone07}. Denote by $p^*_i$ the adjusted
$p$-value for the $i$th window, so that for the level $\alpha_H$, the
corrected $p$-values define set
\[
  \mathfrak B^* = \big\{w_i: p_i^* \leq \alpha_H, i \in [1, N-m+1]\big\}
  \subseteq \mathfrak B.
\] 
The inclusion follows because $\widehat p_i \leq p_i^*$. Depending on the
value of $m$, each element of $\mathfrak B^*$ may end on an translocation
event or not. In the latter case, the extra bases starting after the
last translocation event are deleted.  Let $\mathfrak B^\dagger$ be the
remaining connected regions in $\mathfrak B^*$ after
trimming. The set $\mathfrak B^\dagger$ is finally the group of
hotspots. The significance of each element in $\mathfrak B^\dagger$ is
computed by observing its length, say $\ell$, via
(\ref{eqn:THEpvalue}) by taking $\theta = \ell/N$.

The hotspots here were defined by using $\alpha_H =
0.05$. The method based on the scan statistic with window width $w$
(in base pairs) is denoted SS$_w$ hereafter. Although we considered
several widths, we only present results for the cases SS$_{500}$ and 
SS$_{5000}$.

\subsection{A local approach to hotspot detection}\label{sec:NB_HS}
The probabilistic model for the occurrence of translocations described
in section~\ref{subsec:ss} is implicit in previous work made by
\cite{Klein11} and \cite{Jankovic13} while analysing
hostspots. The data consisting of the genome translocation breakpoints
is represented as a Bernoulli process with success probability $p$,
estimated as $a/N$ with $N$ as the genome length and $a$ as the total
number of observed translocation events. Suppose $(x_i)$, $i=1,
\ldots, a$, are the coordinates of the translocations and let $L_i$
for $i=1, \ldots, a-1$ be the number of bases between $x_i+1$ and
$x_{i+1}$ inclusive. The random variable $L_i$ records therefore the
length until the next translocation starting at $x_i+ 1$, namely
$\ell_i = x_{i+1} - x_i - 1$. The independence of the underlying
Bernoulli process implies that $L_i$ is a geometric random variable
with parameter $p$, that is $\Prb\{L_i = \ell_i\} =
(1-p)^{\ell_i-1}p$, $\ell_i \geq 1$.  Small values for
\begin{equation}
  \label{eqn:localmaxgap}
  \Prb\{L_i \leq \ell_i\} = 1 - (1-p)^{\ell_i}
\end{equation}
may thus be used to detect unusual short distances between successive 
translocation events. In this sense, a hotspot can be defined by
concatenating adjacent segments for which $\Prb\{L_i \leq \ell_i\} \leq
\alpha_c$, where $\alpha_c$ is a given significance level specified in
advance. Suppose that a given sequence of adjacent segments of widths
$\ell_i, \ell_{i+1}, \ldots, \ell_{i+r}$ is identified as a
hotspot. Let 
\[
  L_i^r=\sum_{j=0}^r L_{i+j} \text{ and } 
  \ell_i^r = \sum_{j=0}^r \ell_{i+j},
\]
so that the significance of this hotspot may be quantified by the
$p$-value
\[
 \bar p = \Prb\{L_i^r \leq \ell_i^r\}.
\]
This probability is directly available because $L_i^r$ is a negative 
binomial random variable with parameters $r+1$ and $a/N$, that is 
\[
  \bar p =  \sum_{k=r+1}^{\ell_i^r} \Prb\{L_i^r=k\} 
  = \sum_{k=r}^{\ell_i^r} {k \choose
    r}\Big(\frac{a}{N}\Big)^{r+1}\Big(1 -\frac{a}{N}\Big)^{k-r}.
\]
This method was used by \cite{Klein11} and \cite{Jankovic13}
to define a set of potential hotspots by taking $\alpha_c = 0.01$. Any
candidate of this set is then identified as a  hotspot if:\\ 
\indent ({\it i}) it has more than 3 translocation breakpoints,\\
\indent ({\it ii}) it has at least one read from each of the two sides
of the bait,\\
\indent ({\it iii}) at least 10\% of the translocations come from each
side of the bait,\\
\indent({\it iv}) $\bar p \leq 1\times 10^{-9}$.

Hereafter, we refer to this procedure as the local method and denote
it by NB$_{\alpha_c}$. We describe results obtained with NB$_{0.01}$
and NB$_{0.05}$.

\subsection{Software Implementation}
The genome-wide scan for hotspots according to the method
described in Section~\ref{sec:SS_HS} is implemented by a program we
call \hotscan{}. \hotscan{} is written in Perl and R, and depends 
on the Perl modules Parallel::ForkManager and Math::GSL::SF, available
via CPAN search (\href{http://search.cpan.org}{\tt
  http://search.cpan.org}). The former is required 
for simple parallelisation and the latter to evaluate the
hipergeometric function in (\ref{eqn:THEpvalue}).

\subsection{TC-Seq and ChIP libraries}\label{sec:data}
The TC-Seq data sets analysed here are those described by
\cite{Klein11} and \cite{Jankovic13}. These are deposited at SRA
(\href{http://www.ncbi.nlm.nih.gov/sra}{\tt
  http://www.ncbi.nlm.nih.gov/sra}) under accession numbers  
SRA061477 and SRA039959. These data sets are from four different
translocation libraries: 1. a library from activated B cells infected
with AID expressing retrovirus (denoted hereafter as $\Arv$), 2. a
library from AID deficient B cells (denoted as $\Amm$), 3. a library
from 53BP1 deficient B cells infected with AID retrovirus (denoted as
$\BArv$) and 4. a library from 53BP1 deficient AID deficient B cells
(denoted as $\BAmm$).  The list with the translocation breakpoints
passed on to \hotscan{} in BED format was generated by mapping onto
the reference genome as described in \cite{Klein11}.

The association between translocations hotspots and RNA polymerase II
({\em PolII}) accumulation was examined by using ChIP-seq experiments
deposited at the Gene Expression Omnibus database
\href{http://www.ncbi.nlm.nih.gov/geo}{\tt
  http://www.ncbi.nlm.nih.gov/geo} under the accession number GSE24178.

\subsection{Enrichment analysis}
The analysis of genes targeted by AID that are discovered by
\hotscan{} and NB methods was made by using {\it WebGestalt}
(\cite{Wang13}). The set of genes targeted by AID was compared with
the mouse genome using the hypergeometric test followed by correction
for multiple testing using the Benjamini \& Hochberg method at a
significance level of Top10. The high level functional classification
was based upon \emph{GO Slim} for all three major GO term categories,
namely biological process, cellular component and molecular function.

\section{Results}
\subsection{Scan statistic \& local method}
\begin{figure}
 \centerline{\includegraphics[width=11cm]{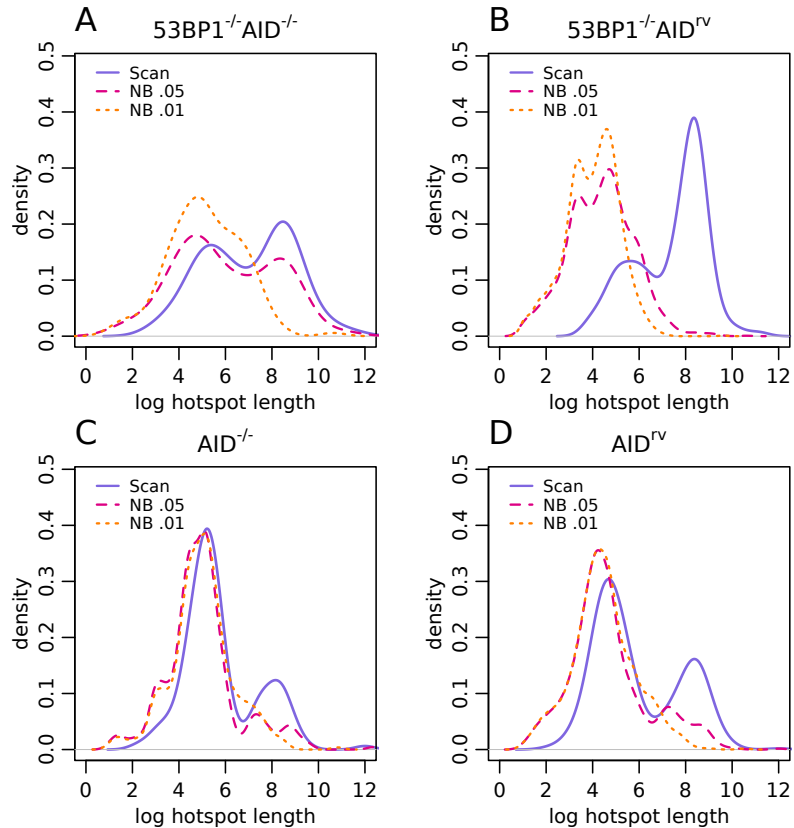}}
 \caption{Hotspot length distributions. A presents the distributions 
  for the $\BAmm$ sample, B for $\BArv$,  C for $\Amm$ and D for
  $\Arv$. The distributions for the hotspots determined with
  SS$_{5000}$ are shown in blue whereas  those for NB$_{0.05}$ and
  NB$_{0.01}$ respectively in magenta and orange. Each graph 
  corresponds to a (Gaussian) kernel density estimate of the 
  underlying distribution.}\label{fig:HS_lengths}
\end{figure}

The methods described in sections \ref{sec:SS_HS} and \ref{sec:NB_HS}
are compared by plotting the distribution of the hotspot lengths
defined by each. Because the observed hotspot lengths vary across
several orders, we considered the logarithm of their actual
length. The results obtained by analysing the four data sets described
in section~\ref{sec:data} are presented in
Figure~\ref{fig:HS_lengths}. The analysis done with SS$_{5000}$ shows 
that the hotspot length distribution is roughly characterised by two
components, one with a mean length of $167 = \lfloor e^{5.12} \rfloor$
base pairs and the other with mean length equal to $4154 = \lfloor
e^{8.332} \rfloor$. Table S1 presents the means, variances and the
weights of these components. Let $0 < \gamma <1$ and $1-\gamma$ be the
weight of the short and long hotspots components respectively. While
the mean position of these components remains almost the same, the
relative weight of long to short hotspots, $\rho = \gamma/(1-\gamma)$,
does shows significant changes. A comparison of the $\Arv$ data set
(Figure~\ref{fig:HS_lengths}.D) and the $\BArv$ data set
(Figure~\ref{fig:HS_lengths}.B) reveals that the relative frequency of
long hotspots is higher in the absence of 53BP1. Indeed, the value for
$\rho$ in the $\BArv$ sample is 1.95, while for the $\Arv$ sample it
is four times smaller, namely $\rho=0.51$. We conclude that 53BP1
decreases the proportion of long hotspots. A similar effect of 53BP1
deficiency is observed in the absence of AID. Indeed, the $\BAmm$
sample (Figure~\ref{fig:HS_lengths}.A) is characterised by $\rho=1.08$
while the $\Amm$ sample (Figure~\ref{fig:HS_lengths}.C) by
$\rho=0.3$. We conclude that in the absence of 53BP1 longer hotspots
are more frequent regardless of AID expression. This effect can be
attributed to the role of 53BP1 in DNA end protection. In the absence
of this protein DNA end resection is increased resulting in longer
hotspots as suggested previously \cite{Jankovic13}. A comparison of
the plots that present the $\Amm$ and $\Arv$ samples
(Figures~\ref{fig:HS_lengths}.C, \ref{fig:HS_lengths}.D respectively)
shows no substantial changes in the proportion of short to long
hotspots, with $\rho=0.3$ and $\rho=0.5$. However, in the absence of
53BP1 the frequency of longer hotspots increases significantly when
AID is overexpressed (Figures~\ref{fig:HS_lengths}.A,
\ref{fig:HS_lengths}.B). Thus, proper DNA repair that is dependent on
53BP1 ensures the predominance of short hotspots, evend when AID is
overexpressed.

Most of these results are not observed when analysing the same data by
the local method described in Section~\ref{sec:NB_HS}. This becomes
clear by inspection of the dashed lines in
Figure~\ref{fig:HS_lengths}, which correspond to the length
distributions for hotspots detected by NB$_{0.01}$ and
NB$_{0.05}$. Even at $\alpha_c= 0.05$, for which one would expect
longer hotspots, the local method is unable to detect the changes in
the frequency of the long hotspots to the extent brought by
\hotscan{}.  It is important to note that by following
(\ref{eqn:localmaxgap}), the local method would classify two
consecutive breakpoints as being part of a hotspot if their distance
is smaller than $\ln(1-\alpha_c)/\ln(1-p)$. Using larger values for
$\alpha_c$ allows thus for larger gaps. Values above 0.05 would
however correspond to tests with Type I Errors higher to what is
commonly acceptable.

% distributions for HS length

\begin{figure*}
  \centering
  \includegraphics[width=15.5cm]{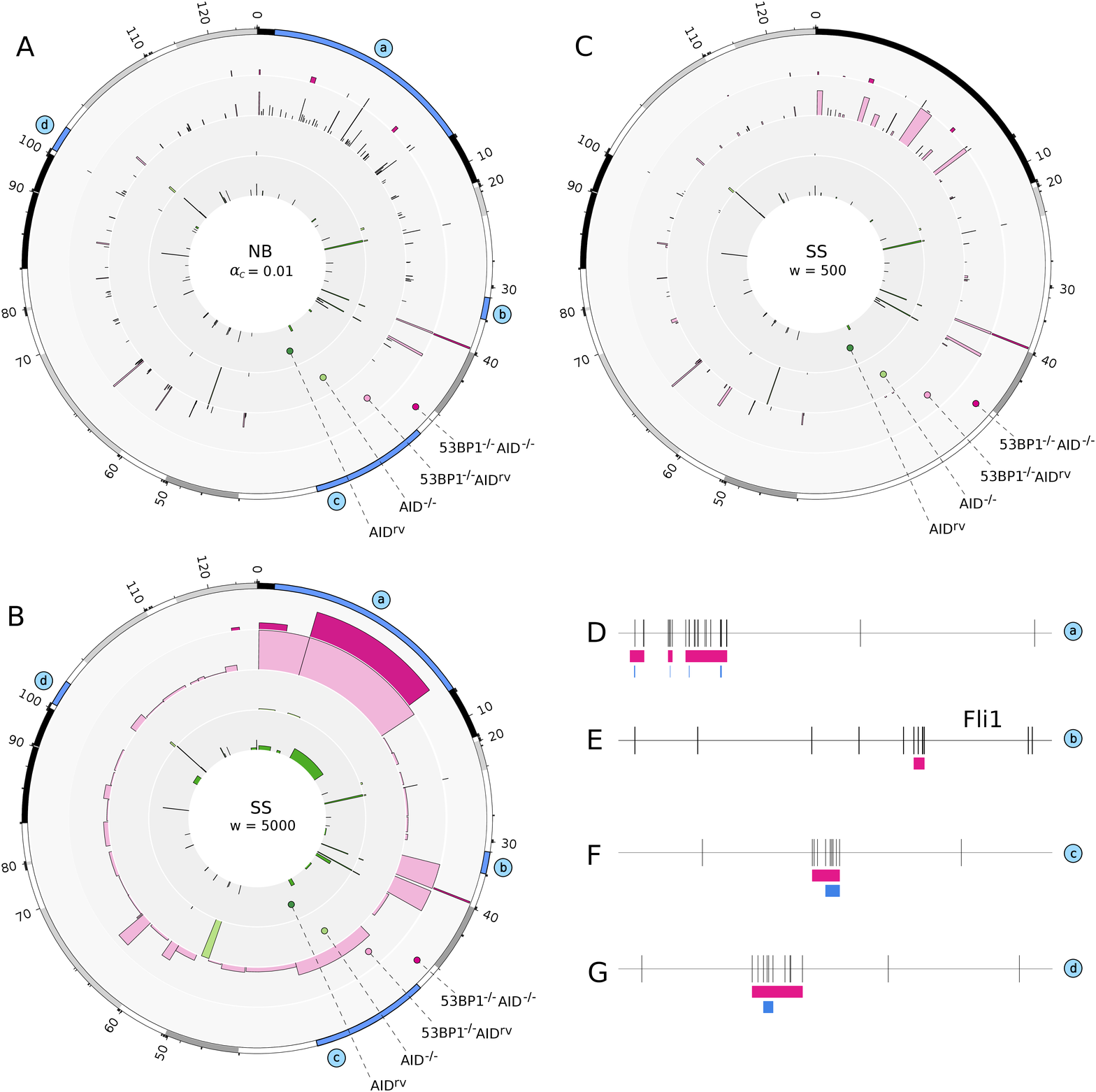}
  \caption{
  Relative position for the hotspots in chromosome 9. The circular
  plots in A, C and B show respectively the hotspots inferred by the
  local method, the scan statistic with $w = 500$ and the scan
  statistic with $w = 5000$. The innermost track presents the hotspots
  for the $\Arv$ data, the next track shows those for the $\Amm$
  sample, then those for $\BArv$ and the outermost track for
  $\BAmm$. Traces in D-G present few examples of the actual
  translocation breakpoints and the estimated hotspots by both
  techniques for the $\Arv$ data. The hotspots by the scan statistic
  (with $w=5000$) are shown in pink and those by the local method in
  blue. The actual locations of these segments within the chromosome
  are identified with the labels (a) - (d) in A and B. The segment in
  E corresponds to the full extension of the \emph{Fli1}
  gene. Circular plots were made with circos \cite{Krzywinski}, by
  plotting the significance of each hotspot along the radial axis as
  $-\log(\bar p)$ in A and $-\log(\hat p)$ in B and C. Due to the
  sparsity of the translocation hotspots, all the hotspot regions in B
  where expanded by using a scaling factor equal to 10000. The same
  regions defined in B where also expanded in A and C to allow for
  comparisons.}\label{fig:circos}
\end{figure*}

The results in Figure~\ref{fig:HS_lengths} present the differences in
the hotspot lengths defined by the scan statistic and the local
method. However, they do not provide any information about the
relative positions of the hotspots detected by either technique. To
address this aspect, we analysed the relative hotspot positions for
all four data sets described in Section~\ref{sec:data}. As an example
Figure~\ref{fig:circos} presents the hotspots for chromosome 9
estimated via $NB_{0.01}$, SS$_{500}$ and SS$_{5000}$. A comparison of
the hotspots found by NB$_{0.01}$ (Figure~\ref{fig:circos}.A) and
those by the scan statistic with a relatively small window, namely by
SS$_{500}$ (Figure~\ref{fig:circos}.C), shows that the hotspots
defined by either method share the same location. However, the
analysis with SS$_{5000}$ (Figure~\ref{fig:circos}.C) reveals the
existence of longer hotspots which include one or few smaller hotspots
found by NB$_{0.01}$. The merging of several smaller hotspots into a
larger one is justified by the sparsity of the data which only becomes
apparent at larger scales. These features are clearly overseen by NB
method because of its local nature. Few examples of the scaling effect
are shown by the examples in Figures~\ref{fig:circos}.D,
\ref{fig:circos}.F and \ref{fig:circos}.G. These results are
consistent with those observed for other chromosomes (see Figures
S1-S2).

Over expression of AID in the absence of 53BP1 results not only in the
increase of the number of translocation but also defines larger
regions where these events cluster. In addition AID overexpression in
$\Bmm$ cells results in a elongation of pre-existing hotspot
regions. This is apparent when comparing the outermost track and the
neighboring one on the circular graph that corresponds to the analysis
with SS$_{5000}$ (Figure~\ref{fig:circos}.C) for the $\BAmm$ and the
$\BArv$ data. The analysis of the same data with a larger scanning
window, namely with SS$_{5000}$, gives a similar result but the
afected regions are much larger (Figure~\ref{fig:circos}.B). The
length of most hotspots in this situation is greatly reduced in 53BP1
suficient samples. This is apparent for the hotspots from the
$\BArv$ and the $\Arv$ data. Interestingly, the effect of 53BP1
correlates with the significance of the hotspots.

We conclude that hotspot length is dependent on 53BP1 and that AID
overexpression in the absence of 53BP1 results in translocations that
cluster over large regions. 

% circos 1: NB & SS

\subsection{Exclusive hotspots}
Most of the more prominent hotspots are defined by both the scan
statistic and the local method (Figures~\ref{fig:circos}.A, C and
Figures S1.A, C and S2.A, C). However, both methods reveal 
exclusive clustering regions (Tables S2, S3). In order to identify
AID-dependent hotspots that are exclusive to each method, we compared
the $\Arv$ and the $\Amm$ data. We found 36 
exclusive hotspots with \hotscan{} (see Table S2) and 27 exclusive
hotspots with NB$_{0.05}$ and NB$_{< 0.01}$ (see Table S3). The
exclusive hotspots obtained by the scan statistic were defined using
different  window widths (50, 100, 150, 250, 500, 1000, 2500,
and 5000 bp). Regions that are identified as exclusive AID hotspots
were also analysed for several biologically relevant markers. First,
we analysed whether our exclusive hotspots correlate with Replication
Protein A (RPA) binding sites in activated B cells \cite{Yamane11,
Hakim12}. The sites of RPA accumulation have been shown to overlap
well with AID targets genome-wide and it was proposed that RPA marks
AID induced DNA double strand breaks. Further, we analysed the overlap
with sites where RNA Polymerase II (\emph{PolII}) accumulates as it
was shown that transcription is necessary for AID targeting
(\cite{DiNoia07}). We also analysed the overlap with known fragile
regions, namely  by the Early Replicating Fragile Sites (ERFS) and
Common Fragile Sites (CFS) \cite{Barlow13}. The results of all
these comparisons are summarised in Tables S2 and S3. A total of 20 of
the 36 (55.5\%) exclusive hotspots found by the scan statistic were
common to all sites. Notably, all of these sites are associated with
the \emph{PolII} signal (Table S2). On the other hand, only 8 of the
27 (29.6\%) exclusive hotspots of the local method fall within these
sites and 6 are associated with the \emph{PolII} signal (Table
S3). Thus, the hotspots defined by the scan statistic show higher
correlation with active transcription, RPA accumulation and common
fragile sites than those defined by the local method.

AID leads to the accumulation of somatic mutations in a large number
of non immunoglobulin genes \cite{ZweiNusse10}. A analysis for the
presence of somatic hypermutations in 1,496,058 bp from activated B
cells \cite{Yamane11} revealed a number of non-immunoglobulin genes
with AID dependent mutations: \emph{Il4ra}, \emph{Grap},
\emph{Hist1h1c}, \emph{Ly6e}, \emph{Gadd45g} and \emph{Il4i1}. Three
of these, namely \emph{Il4ra}, \emph{Grap} and \emph{Ly6e}, were
detected as genes with AID dependent hotspots by both methods, but a
hotspot in \emph{Hist1h1c} (mutation rate in \emph{Ig$\kappa$-AID}
\emph{Ung}$^{-/-}$: 79.7 $x 10^{-5}$), was only found by \hotscan{}
(Table S2 and Figure S3). Three other genes associated with
chromosomal translocations where detected exclusively by \hotscan{},
namely \emph{Fli1}, \emph{Dlx5} and \emph{Birc3}. The
\emph{Fli1}(Friend leukemia integration 1) gene
(Figure~\ref{fig:circos}.E) is translocated in 90\% of Ewing sarcomas
and is important in tumorigenesis \cite{Riggi07}. \emph{Dlx5}
(distal-less homeobox 5) is implicated in T-Cell Lymphomas
(\cite{Tan08}). Finally, the \emph{Birc3} (baculoviral IAP
repeat-containing 3) gene encodes an apoptosis inhibitor that is
associated with MALT lymphomas \cite{Dierlamm99}. A complementary
enrichment analysis \cite{Wang13} for the genes identified by
\hotscan{} is included in Figures S5, S8 and Tables S4, S5. The
functional categories associated with the scan statistic hotspots
indicate that the top ranked genes are important in B lymphocytes.

\section{Discussion}
Here we describe a method for the identification of chromosomal
translocation hotspots. In contrast to the previous methods, the
significance level for a cluster is defined on a chomosome-wide basis
by using scan statistics. We show that this has important consequences
in the analysis of translocation hotspots in primary B cells in the
prescence or absence of 53BP1 repair protein.

The previous study by \cite{Jankovic13} showed that 53BP1 deficiency
results in an increase of rearrangements to intergenic regions and
changes the frequency and distribution of translocations in
$\gamma_3$, $\gamma_1$ immunoglobulin switch regions and other 16
prominent hotspots.  Our analysis adds to these findings by showing
that the 53BP1 deficiency results in the overal enrichment of longer 
hotspots. These results support the previous conclusion that 53BP1
prevents the resection of DNA thus resulting in shorter hotspots 
\cite{Jankovic13}. Our analysis here also shows that an increased
amount of AID results in quite a substantial enlargement of 
pre-existing hotspot  regions. These changes can only be observed with
wider scanning windows, here $w=5000$, and are not detected by
previous methods because of their local characterization of
clustering. The success of the scan statistic here is brought by its
ability to detect events spread across several scales as is shown by
the analyses made with several scanning window widths. Our analysis
with the scan statistic is able to identify several exclusive hotspots
whose authenticity is supported by independent experimental
approaches. Some  of these exclusive events are localised in genes
that are known to be relevant in tumorigenesis.

The approach presented here may be applied to a variety of questions
related to the detection of unusual clustering of a given pattern
throughout the genome. Few recent examples of particular interest are
the detection of enriched genomic interaction regions such as those
defined via ChIP-seq experiments \cite{Ma11}, 4C-seq experiments
\cite{Simonis06} and DNA-DNA contact sites \cite{deWit12}. We expect
our method to be especially useful for the analysis of data where a
global significance to clustering can be considered.

\section*{Acknoledgement}
I. T. S. wishes to thank T. Oliveira for kindly providing the
script for the local method described in section~\ref{sec:NB_HS}, 
R. A. R. thanks K. J. Abraham for useful discussions. M. C. N. is a
Howard Huges Medical Institute Inverstigator.

\end{document}